\documentclass[twocolumn]{aastex631}

\usepackage{graphicx}
\usepackage{float}
\usepackage{txfonts}
\usepackage{soul}
\usepackage{ulem}
\usepackage{booktabs}
\usepackage{amsfonts}
\usepackage{enumitem}
\usepackage{listings}
\setlist{nolistsep}
\newcommand{\PSF}{PSF}

\newcommand{\DESDRLINK}{\url{https://github.com/des-science/DES-SN5YR}}
\newcommand{\DESSN}{DES-SN}
\newcommand{\DESYt}{DES-SN3YR}
\newcommand{\DESYf}{DES-SN5YR}
\newcommand{\DIA}{\texttt{DiffImg}}
\newcommand{\SMP}{SMP}
\newcommand{\NLCDIFFIMG}{$31,636$}
\newcommand{\NLCSMP}{$19,706$}
\newcommand{\NCosmoSNe}{$1635$}

\usepackage{appendix}
\usepackage{lipsum}  

\reportnum{DES-2023-0807}
\reportnum{FERMILAB-PUB-24-0292-PPD} 

\shorttitle{DES SN Y5: SMP}
\shortauthors{DES Collaboration}

\graphicspath{{./}}

\begin{document}

\title{The Dark Energy Survey Supernova Program: Light curves and 5-Year data release}

\submitjournal{The Astrophysical Journal}



\author[0000-0002-8687-0669]{B.~O.~S\'anchez}
\affiliation{Aix Marseille Univ, CNRS/IN2P3, CPPM, Marseille, France}
\affiliation{Department of Physics, Duke University Durham, NC 27708, USA}

\author[0000-0001-5201-8374]{D.~Brout}
\affiliation{Departments of Astronomy and Physics, Boston University, Boston, MA 02140, USA}

\author{M.~Vincenzi}
\affiliation{Department of Physics, Duke University Durham, NC 27708, USA}

\author{M.~Sako}
\affiliation{Department of Physics and Astronomy, University of Pennsylvania, Philadelphia, PA 19104, USA}

\author[0000-0001-6718-2978]{K.~Herner}
\affiliation{Fermi National Accelerator Laboratory, P. O. Box 500, Batavia, IL 60510, USA}


\author[0000-0003-3221-0419]{R.~Kessler}
\affiliation{Department of Astronomy and Astrophysics, University of Chicago, Chicago, IL 60637, USA}
\affiliation{Kavli Institute for Cosmological Physics, University of Chicago, Chicago, IL 60637, USA}

\author[0000-0002-4213-8783]{T.~M.~Davis}
\affiliation{School of Mathematics and Physics, University of Queensland,  Brisbane, QLD 4072, Australia}

\author{D.~Scolnic}
\affiliation{Department of Physics, Duke University Durham, NC 27708, USA}

\author[0000-0002-5389-7961]{M.~Acevedo}
\affiliation{Department of Physics, Duke University Durham, NC 27708, USA}

\author[0000-0001-6633-9793]{J.~Lee}
\affiliation{Department of Physics and Astronomy, University of Pennsylvania, Philadelphia, PA 19104, USA}

\author{A.~M\"oller}
\affiliation{Centre for Astrophysics \& Supercomputing, Swinburne University of Technology, Victoria 3122, Australia}

\author{H.~Qu}
\affiliation{Department of Physics and Astronomy, University of Pennsylvania, Philadelphia, PA 19104, USA}

\author{L.~Kelsey}
\affiliation{Institute of Cosmology and Gravitation, University of Portsmouth, Portsmouth, PO1 3FX, UK}
\affiliation{School of Physics and Astronomy, University of Southampton,  Southampton, SO17 1BJ, UK}

\author{P.~Wiseman}
\affiliation{School of Physics and Astronomy, University of Southampton,  Southampton, SO17 1BJ, UK}


\author[0000-0003-1997-3649]{P.~Armstrong}
\affiliation{The Research School of Astronomy and Astrophysics, Australian National University, ACT 2601, Australia}

\author[0000-0002-1873-8973]{B.~Rose}
\affiliation{Department of Physics, Baylor University, One Bear Place \#97316, Waco, TX 76798-7316, USA}
\affiliation{Department of Physics, Duke University Durham, NC 27708, USA}

\author{R.~Camilleri}
\affiliation{School of Mathematics and Physics, University of Queensland,  Brisbane, QLD 4072, Australia}

\author{R.~Chen}
\affiliation{Department of Physics, Duke University Durham, NC 27708, USA}

\author{L.~Galbany}
\affiliation{Institut d'Estudis Espacials de Catalunya (IEEC), 08034 Barcelona, Spain}
\affiliation{Institute of Space Sciences (ICE, CSIC),  Campus UAB, Carrer de Can Magrans, s/n,  08193 Barcelona, Spain}

\author{E.~Kovacs}
\affiliation{Argonne National Laboratory, 9700 South Cass Avenue, Lemont, IL 60439, USA}

\author[0000-0003-1731-0497]{C.~Lidman}
\affiliation{Centre for Gravitational Astrophysics, College of Science, The Australian National University, ACT 2601, Australia}
\affiliation{The Research School of Astronomy and Astrophysics, Australian National University, ACT 2601, Australia}

\author[0000-0002-8012-6978]{B.~Popovic}
\affiliation{Univ Lyon, Univ Claude Bernard Lyon 1, CNRS, IP2I Lyon / IN2P3, IMR 5822, F-69622, Villeurbanne, France}
\affiliation{Department of Physics, Duke University Durham, NC 27708, USA}

\author[0000-0002-3321-1432]{M.~Smith}
\affiliation{Physics Department, Lancaster University, Lancaster, LA1 4YB, UK}

\author[0000-0001-9053-4820]{M.~Sullivan}
\affiliation{School of Physics and Astronomy, University of Southampton,  Southampton, SO17 1BJ, UK}

\author[0000-0001-6882-0230]{M.~Toy}
\affiliation{School of Physics and Astronomy, University of Southampton,  Southampton, SO17 1BJ, UK}

\author{D.~Carollo}
\affiliation{INAF-Osservatorio Astronomico di Trieste, via G. B. Tiepolo 11, I-34143 Trieste, Italy}

\author{K.~Glazebrook}
\affiliation{Centre for Astrophysics \& Supercomputing, Swinburne University of Technology, Victoria 3122, Australia}

\author[0000-0003-3081-9319]{G.~F.~Lewis}
\affiliation{Sydney Institute for Astronomy, School of Physics, A28, The University of Sydney, NSW 2006, Australia}

\author{R.~C.~Nichol}
\affiliation{School of Mathematics and Physics, University of Surrey, Guildford, Surrey, GU2 7XH, UK}

\author{B.~E.~Tucker}
\affiliation{The Research School of Astronomy and Astrophysics, Australian National University, ACT 2601, Australia}


\author{T.~M.~C.~Abbott}
\affiliation{Cerro Tololo Inter-American Observatory, NSF's National Optical-Infrared Astronomy Research Laboratory, Casilla 603, La Serena, Chile}

\author{M.~Aguena}
\affiliation{Laborat\'orio Interinstitucional de e-Astronomia - LIneA, Rua Gal. Jos\'e Cristino 77, Rio de Janeiro, RJ - 20921-400, Brazil}

\author[0000-0002-7069-7857]{S.~Allam}
\affiliation{Fermi National Accelerator Laboratory, P. O. Box 500, Batavia, IL 60510, USA}

\author{O.~Alves}
\affiliation{Department of Physics, University of Michigan, Ann Arbor, MI 48109, USA}

\author[0000-0002-0609-3987]{J.~Annis}
\affiliation{Fermi National Accelerator Laboratory, P. O. Box 500, Batavia, IL 60510, USA}

\author{J.~Asorey}
\affiliation{Departamento de Física Teórica and IPARCOS, Universidad Complutense de Madrid, 28040 Madrid, Spain}

\author{S.~Avila}
\affiliation{Institut de F\'{\i}sica d'Altes Energies (IFAE), The Barcelona Institute of Science and Technology, Campus UAB, 08193 Bellaterra (Barcelona) Spain}

\author{D.~Bacon}
\affiliation{Institute of Cosmology and Gravitation, University of Portsmouth, Portsmouth, PO1 3FX, UK}

\author[0000-0002-8458-5047]{D.~Brooks}
\affiliation{Department of Physics \& Astronomy, University College London, Gower Street, London, WC1E 6BT, UK}

\author{D.~L.~Burke}
\affiliation{Kavli Institute for Particle Astrophysics \& Cosmology, P. O. Box 2450, Stanford University, Stanford, CA 94305, USA}
\affiliation{SLAC National Accelerator Laboratory, Menlo Park, CA 94025, USA}

\author[0000-0003-3044-5150]{A.~Carnero~Rosell}
\affiliation{Instituto de Astrofisica de Canarias, E-38205 La Laguna, Tenerife, Spain}
\affiliation{Laborat\'orio Interinstitucional de e-Astronomia - LIneA, Rua Gal. Jos\'e Cristino 77, Rio de Janeiro, RJ - 20921-400, Brazil}

\author[0000-0002-3130-0204]{J.~Carretero}
\affiliation{Institut de F\'{\i}sica d'Altes Energies (IFAE), The Barcelona Institute of Science and Technology, Campus UAB, 08193 Bellaterra (Barcelona) Spain}

\author[0000-0001-7316-4573]{F.~J.~Castander}
\affiliation{Institut d'Estudis Espacials de Catalunya (IEEC), 08034 Barcelona, Spain}
\affiliation{Institute of Space Sciences (ICE, CSIC),  Campus UAB, Carrer de Can Magrans, s/n,  08193 Barcelona, Spain}

\author{L.~N.~da Costa}
\affiliation{Laborat\'orio Interinstitucional de e-Astronomia - LIneA, Rua Gal. Jos\'e Cristino 77, Rio de Janeiro, RJ - 20921-400, Brazil}

\author{J.~Duarte}
\affiliation{CENTRA, Instituto Superior T\'ecnico, Universidade de Lisboa, Av. Rovisco Pais 1, 1049-001 Lisboa, Portugal}

\author{M.~E.~S.~Pereira}
\affiliation{Hamburger Sternwarte, Universit\"{a}t Hamburg, Gojenbergsweg 112, 21029 Hamburg, Germany}

\author[0000-0002-0466-3288]{S.~Desai}
\affiliation{Department of Physics, IIT Hyderabad, Kandi, Telangana 502285, India}

\author[0000-0002-8357-7467]{H.~T.~Diehl}
\affiliation{Fermi National Accelerator Laboratory, P. O. Box 500, Batavia, IL 60510, USA}

\author{S.~Everett}
\affiliation{Jet Propulsion Laboratory, California Institute of Technology, 4800 Oak Grove Dr., Pasadena, CA 91109, USA}

\author{I.~Ferrero}
\affiliation{Institute of Theoretical Astrophysics, University of Oslo. P.O. Box 1029 Blindern, NO-0315 Oslo, Norway}

\author[0000-0002-2367-5049]{B.~Flaugher}
\affiliation{Fermi National Accelerator Laboratory, P. O. Box 500, Batavia, IL 60510, USA}

\author[0000-0003-4079-3263]{J.~Frieman}
\affiliation{Fermi National Accelerator Laboratory, P. O. Box 500, Batavia, IL 60510, USA}
\affiliation{Kavli Institute for Cosmological Physics, University of Chicago, Chicago, IL 60637, USA}

\author[0000-0002-9370-8360]{J.~Garc\'ia-Bellido}
\affiliation{Instituto de Fisica Teorica UAM/CSIC, Universidad Autonoma de Madrid, 28049 Madrid, Spain}

\author{M.~Gatti}
\affiliation{Department of Physics and Astronomy, University of Pennsylvania, Philadelphia, PA 19104, USA}

\author[0000-0001-9632-0815]{E.~Gaztanaga}
\affiliation{Institut d'Estudis Espacials de Catalunya (IEEC), 08034 Barcelona, Spain}
\affiliation{Institute of Cosmology and Gravitation, University of Portsmouth, Portsmouth, PO1 3FX, UK}
\affiliation{Institute of Space Sciences (ICE, CSIC),  Campus UAB, Carrer de Can Magrans, s/n,  08193 Barcelona, Spain}

\author[0000-0002-3730-1750]{G.~Giannini}
\affiliation{Institut de F\'{\i}sica d'Altes Energies (IFAE), The Barcelona Institute of Science and Technology, Campus UAB, 08193 Bellaterra (Barcelona) Spain}
\affiliation{Kavli Institute for Cosmological Physics, University of Chicago, Chicago, IL 60637, USA}

\author{S. Gonz\'alez-Gait\'an}
\affiliation{CENTRA, Instituto Superior T\'ecnico, Universidade de Lisboa, Av. Rovisco Pais 1, 1049-001 Lisboa, Portugal}

\author{R.~A.~Gruendl}
\affiliation{Center for Astrophysical Surveys, National Center for Supercomputing Applications, 1205 West Clark St., Urbana, IL 61801, USA}
\affiliation{Department of Astronomy, University of Illinois at Urbana-Champaign, 1002 W. Green Street, Urbana, IL 61801, USA}

\author[0000-0003-0825-0517]{G.~Gutierrez}
\affiliation{Fermi National Accelerator Laboratory, P. O. Box 500, Batavia, IL 60510, USA}

\author{S.~R.~Hinton}
\affiliation{School of Mathematics and Physics, University of Queensland,  Brisbane, QLD 4072, Australia}

\author{D.~L.~Hollowood}
\affiliation{Santa Cruz Institute for Particle Physics, Santa Cruz, CA 95064, USA}

\author[0000-0002-6550-2023]{K.~Honscheid}
\affiliation{Center for Cosmology and Astro-Particle Physics, The Ohio State University, Columbus, OH 43210, USA}
\affiliation{Department of Physics, The Ohio State University, Columbus, OH 43210, USA}

\author[0000-0001-5160-4486]{D.~J.~James}
\affiliation{Center for Astrophysics $\vert$ Harvard \& Smithsonian, 60 Garden Street, Cambridge, MA 02138, USA}

\author[0000-0003-0120-0808]{K.~Kuehn}
\affiliation{Australian Astronomical Optics, Macquarie University, North Ryde, NSW 2113, Australia}
\affiliation{Lowell Observatory, 1400 Mars Hill Rd, Flagstaff, AZ 86001, USA}

\author[0000-0002-1134-9035]{O.~Lahav}
\affiliation{Department of Physics \& Astronomy, University College London, Gower Street, London, WC1E 6BT, UK}

\author{S.~Lee}
\affiliation{Jet Propulsion Laboratory, California Institute of Technology, 4800 Oak Grove Dr., Pasadena, CA 91109, USA}

\author[0000-0002-7825-3206]{H.~Lin}
\affiliation{Fermi National Accelerator Laboratory, P. O. Box 500, Batavia, IL 60510, USA}

\author[0000-0003-0710-9474]{J.~L.~Marshall}
\affiliation{George P. and Cynthia Woods Mitchell Institute for Fundamental Physics and Astronomy, and Department of Physics and Astronomy, Texas A\&M University, College Station, TX 77843,  USA}

\author[0000-0001-9497-7266]{J. Mena-Fern{\'a}ndez}
\affiliation{LPSC Grenoble - 53, Avenue des Martyrs 38026 Grenoble, France}

\author[0000-0002-6610-4836]{R.~Miquel}
\affiliation{Instituci\'o Catalana de Recerca i Estudis Avan\c{c}ats, E-08010 Barcelona, Spain}
\affiliation{Institut de F\'{\i}sica d'Altes Energies (IFAE), The Barcelona Institute of Science and Technology, Campus UAB, 08193 Bellaterra (Barcelona) Spain}

\author{J.~Myles}
\affiliation{Department of Astrophysical Sciences, Princeton University, Peyton Hall, Princeton, NJ 08544, USA}

\author[0000-0003-2120-1154]{R.~L.~C.~Ogando}
\affiliation{Observat\'orio Nacional, Rua Gal. Jos\'e Cristino 77, Rio de Janeiro, RJ - 20921-400, Brazil}

\author[0000-0002-6011-0530]{A.~Palmese}
\affiliation{Department of Physics, Carnegie Mellon University, Pittsburgh, Pennsylvania 15312, USA}

\author[0000-0001-9186-6042]{A.~Pieres}
\affiliation{Laborat\'orio Interinstitucional de e-Astronomia - LIneA, Rua Gal. Jos\'e Cristino 77, Rio de Janeiro, RJ - 20921-400, Brazil}
\affiliation{Observat\'orio Nacional, Rua Gal. Jos\'e Cristino 77, Rio de Janeiro, RJ - 20921-400, Brazil}

\author[0000-0002-2598-0514]{A.~A.~Plazas~Malag\'on}
\affiliation{Kavli Institute for Particle Astrophysics \& Cosmology, P. O. Box 2450, Stanford University, Stanford, CA 94305, USA}
\affiliation{SLAC National Accelerator Laboratory, Menlo Park, CA 94025, USA}

\author{A.~Porredon}
\affiliation{Ruhr University Bochum, Faculty of Physics and Astronomy, Astronomical Institute, German Centre for Cosmological Lensing, 44780 Bochum, Germany}

\author[0000-0002-9328-879X]{A.~K.~Romer}
\affiliation{Department of Physics and Astronomy, Pevensey Building, University of Sussex, Brighton, BN1 9QH, UK}

\author[0000-0002-9646-8198]{E.~Sanchez}
\affiliation{Centro de Investigaciones Energ\'eticas, Medioambientales y Tecnol\'ogicas (CIEMAT), Madrid, Spain}

\author[0000-0003-3054-7907]{D.~Sanchez Cid}
\affiliation{Centro de Investigaciones Energ\'eticas, Medioambientales y Tecnol\'ogicas (CIEMAT), Madrid, Spain}

\author[0000-0002-1831-1953]{I.~Sevilla-Noarbe}
\affiliation{Centro de Investigaciones Energ\'eticas, Medioambientales y Tecnol\'ogicas (CIEMAT), Madrid, Spain}

\author[0000-0002-7047-9358]{E.~Suchyta}
\affiliation{Computer Science and Mathematics Division, Oak Ridge National Laboratory, Oak Ridge, TN 37831}

\author{M.~E.~C.~Swanson}
\affiliation{Center for Astrophysical Surveys, National Center for Supercomputing Applications, 1205 West Clark St., Urbana, IL 61801, USA}

\author[0000-0003-1704-0781]{G.~Tarle}
\affiliation{Department of Physics, University of Michigan, Ann Arbor, MI 48109, USA}

\author[0000-0001-7211-5729]{D.~L.~Tucker}
\affiliation{Fermi National Accelerator Laboratory, P. O. Box 500, Batavia, IL 60510, USA}

\author{V.~Vikram}
\affiliation{Argonne National Laboratory, 9700 S Cass Ave, Lemont, IL 60439, USA}

\author[0000-0002-7123-8943]{A.~R.~Walker}
\affiliation{Cerro Tololo Inter-American Observatory, NSF's National Optical-Infrared Astronomy Research Laboratory, Casilla 603, La Serena, Chile}

\author{N.~Weaverdyck}
\affiliation{Department of Astronomy, University of California, Berkeley,  501 Campbell Hall, Berkeley, CA 94720, USA}
\affiliation{Lawrence Berkeley National Laboratory, 1 Cyclotron Road, Berkeley, CA 94720, USA}

\collaboration{100}{(DES Collaboration)}

\correspondingauthor{B.~O.~S\'anchez}
\email{bsanchez@cppm.in2p3.fr}

\begin{abstract} 
We present $griz$ photometric light curves for the full 5 years of the Dark Energy Survey Supernova program (\DESSN{}), obtained with both forced Point Spread Function (\PSF{}) photometry on Difference Images (\DIA{}) performed during survey operations, and Scene Modelling Photometry (\SMP{}) on search images processed after the survey. This release contains \NLCDIFFIMG{} \DIA{} and \NLCSMP{} high-quality \SMP{} light curves, the latter of which contains \NCosmoSNe{} photometrically-classified supernovae that pass cosmology quality cuts. This sample spans the largest redshift ($z$) range ever covered by a single SN survey ($0.1<z<1.13$) and is the largest single sample from a single instrument of SNe ever used for cosmological constraints. We describe in detail the improvements made to obtain the final \DESSN{} photometry and provide a comparison to what was used in the \DESYt{} spectroscopically-confirmed SN Ia sample. We also include a comparative analysis of the performance of the \SMP{} photometry with respect to the real-time \DIA{} forced photometry and find that \SMP{} photometry is more precise, more accurate, and less sensitive to the host-galaxy surface brightness anomaly. The public release of the light curves and ancillary data can be found at \DESDRLINK{}. Finally, we discuss implications for future transient surveys, such as the forthcoming Vera Rubin Observatory Legacy Survey of Space and Time (LSST).
\end{abstract}


\section{Introduction} \label{sec:intro}
Type Ia Supernovae (SNe Ia) are an established cosmological probe, used to discover the accelerating expansion of the universe \citep{riess98,perlmutter99}, and constrain the dark energy equation of state parameter, $w$. The Dark Energy Survey, conceived in the period following the discovery of the accelerating universe, has completed a 5-year SN Ia discovery and follow-up program (\DESSN{}) using repeated (1-week cadence) observations of 27 square degrees with the Dark Energy Camera \citep[DECam][]{decam_flaugher_2015} at Cerro Tololo International Observatory starting in August 2013 and ending in January 2018. It has assembled the largest sample of SNe~Ia ever observed with a single instrument with \NCosmoSNe{} photometrically classified SNe Ia suitable for cosmology over the redshift interval $0.1<z<1.13$.

A first set of cosmology results using only the first 3 years (\DESYt{}) of spectroscopically confirmed SN Ia was released in 2019 \citep{brout19ana,des3yr,dandrea_dessn_2018,Brout19smp}, including a total of 251 light-curves. The purpose of this first analysis was to provide competitive constraints, to spur the development of key data processing and analysis pipelines, and also to identify areas for improvement. The tools for the 5-year analysis were developed over many years and are reported in a number of papers. These software tools include 
  difference-imaging \citep[\DIA{}]{Kessler2015}, 
  scene-model photometry \citep[\SMP{}]{Brout19smp}, 
  chromatic corrections \citep{burke_fgcmdes_2018,lasker18}, 
  simulated bias corrections \citep[K19]{kessler19}, 
  host-mass correlations \citep{Smith_2020, kelsey_environment_2021}, 
  photometric classification \citep{vincenzi_spectro_2019, moller_des_snn_2022, moller_pc_2020}, 
  and a comprehensive analysis framework (\citealp[SNANA K09]{kessler_snana_2009}; \citealp[Pippin]{pippin}; \citealp{bhs21};\citealp{kva23}; \citealp{qu2024dark}; \citealp{Armstrong_2023}). 
The complete analysis using these software tools is presented in \citet{vincenzi_DES5Ysyst_2024}; and \citet{DES_SN5Y_cosmo_2024}.

For the 5-year analysis, DES has compiled a sample of photometrically classified SNe Ia that is an order of magnitude larger than that used in the spectroscopically classified \DESYt{} analysis. This larger sample, hereafter \DESYf{}, has resulted in a 30\% reduction in uncertainties in the dark energy equation of state parameter, $w$, with respect to the initial \DESYt{} analysis \citep{brout19ana}. The uncertainty reduction is less than the naive $\sqrt{N_{\rm SNe}}$ because 
   i) the number of spec-confirmed low-redshift events is similar to that in the 3-year analysis, 
   ii) the CMB constraint is only slightly improved w.r.t. 3-year analysis, and 
   iii) the additional events with photometric classification tend to have lower signal-to-noise ratio compared to the spec-confirmed sample.
However, along with smaller statistical uncertainties comes an increased need to reduce systematic uncertainties. Additionally, as light curves are used for the classification itself there is also a need to improve the photometry. Classification was performed using SuperNNova \citep[][]{moller_pc_2020},\footnote{\url{https://github.com/supernnova/SuperNNova}} using a simulated training set that includes multiple non-Ia supernova types to capture diverse sources of contamination \citep{moller_des_snn_2022,vincenzi2023classification}. 

This work focuses on the extraction of the photometric light-curve fluxes from the five year observations of the DES Supernova Program, and the subsequent public data release. Past cosmological analyses have long utilized two main methods for extracting supernova photometry: Difference Imaging followed by point-spread-function (\PSF{}) fitting photometry on co-added difference images (hereafter \DIA{}), and a forward modeling method called Scene Modeling Photometry (\SMP{}). In this paper, we present the photometry measurements using both methods. The \DIA{} photometry was produced during DES operations with the primary purpose of discovering transients. \SMP{} was run independently of DES operations, with the purpose of high-precision photometry for the cosmology analysis; only a subset of single-season \DIA{} candidates were able to be processed because \SMP{} is not designed for multi-season light curves (e.g.\ AGN).
We also provide details on the ancillary data included as part of this data release.

The structure of the paper is as follows. 
In Sec.~2 we describe the photometric calibration of the images and the different photometric methods (\DIA{} and \SMP{}) applied to discovered transients in them. We also provide consistency checks, and describe improvements. 
In Sec.~3 we summarize the auxiliary data (redshifts, host-galaxy parameters and photometry) provided in this release.
In Sec.~4 we provide some concluding remarks and in Sec.~5 we describe the format of the Data Release.

\section{Data}\label{sec:calibration}
\subsection{DES Supernova Program Overview}\label{sec:desprogram}
\begin{figure*}
    \centering
    \includegraphics[width=\textwidth]{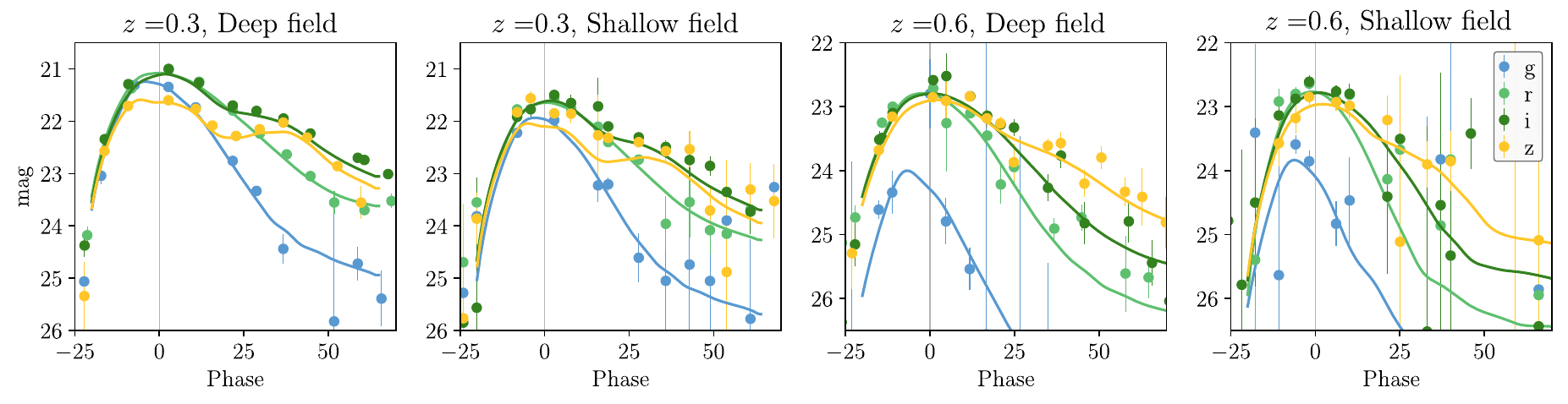}
    \caption{DES light curves of SNe at different redshifts ($z\sim0.3$ and $z\sim0.6$) in the SN Deep and SN Shallow fields. Lines are the SALT3 model that best fits the light curve data.}
    \label{fig:LC}
\end{figure*}

The \DESSN{} program \citep{dandrea_dessn_2018} ran from 2013 through 2018, spanning 5 years and obtained images of 10 fields, (8 shallow and 2 deep fields) in a combined footprint of $\sim27$~deg$\hbox{}^2$. The observing strategy and typical cadence, seeing and depths are provided in \cite{dandrea_dessn_2018} and \cite{neilsen_DESOStrat_2019}. Transients were detected in before the start of the next observing evening using the Difference Imaging technique described in \cite{Kessler2015}, and artifacts were removed using the candidate classification methodology developed in \cite{Goldstein_2015}. An extensive host-galaxy spectroscopic follow-up program was performed using the 2dF fibre positioner and AAOmega spectrograph on the Anglo-Australian Telescope, as part of the OzDES survey \citep{ozdes_yuan_2015,Childress_OzDES_2017, lidman_OzDES_2020}. 
Examples of DES SN light curves at different redshifts in both the \DESSN{} Deep and Shallow fields are presented in Fig.~\ref{fig:LC}.

\subsection{Photometric Calibration}

For SN Ia cosmological measurements, it is essential to both accurately determine the inter-filter calibration within a survey (especially if the survey spans a wide range of redshifts) and the inter-survey calibration (when datasets from multiple surveys are combined).

First, DES images are internally calibrated using a catalog of 17 million tertiary standard stars within the DES footprint built using the Forward Global Calibration Method (FGCM) as conceived by \citet[][]{Stubbs_2006} and implemented in DES by \citet[][]{burke_fgcmdes_2018}. This method provides excellent all-sky uniformity of $<3$ mmag for DES \citep{Sevilla_Noarbe_2021,rykoff23}. 

The FGCM tertiary standard star catalog provided in \citet[][]{burke_fgcmdes_2018} was utilized in the DES-SN3YR cosmological analysis. The FGCM catalog was updated in the period between \DESYt{} and \DESYf{} and here we use the stellar catalogues as presented in Appendix 3 of \cite{Sevilla_Noarbe_2021}. The improvements are summarized as follows: \textit{(i)} improved corrections to aperture photometry, \textit{(ii)} an update to the DES Y3A2 standard bandpasses \citep[see ][Sec. 4.3]{Sevilla_Noarbe_2021}, \textit{(iii)} improved uniformity in years following the bad weather of year 3 \citep{diehl2016dark}, \textit{iv}) improved astrometry using the longer temporal baseline, and \textit{v}) other technical and practical improvements.

SN~Ia cosmology analyses, including \citet{vincenzi_DES5Ysyst_2024}, use multiple surveys to cover both low redshift and high redshift needed for competitive cosmological constraints. For this reason, we utilize the calibration of \citet[][Supercal-Fragilistic]{Fragilistic} which is an improvement over the \citet[][Supercal]{Supercal} method. This method consists of simultaneously cross-calibrating the FGCM catalog with the stellar catalogs from numerous other wide-field surveys (e.g.\ PS1, SDSS, SNLS).  The Supercal-Fragilistic methodology consists in determining a global solution that minimizes the differences between each survey by using their published calibrations as prior information. Supercal-Fragilistic find similar sign of offsets for DES [$+0.002,-0.009,-0.007,+0.006$] in [$g,r,i,z$] as those found in \citet[][]{rykoff23} [$+0.001,-0.003,-0.001,+0.002$], but of larger magnitude; though they find that these offsets are consistent with each other given that the external tertiary cross-calibration data used to perform the calibration in Supercal-Fragilistic is independent from \citet[][]{rykoff23}. 

In this work we have chosen to adopt the offsets from Supercal-Fragilistic because: 
  1) the low-$z$ samples used with the DES-5YR sample to constrain cosmology have been calibrated in Supercal-Fragilistic, 
  2) included is the covariance between DECam filters and low-$z$ filters utilized in the cosmology likelihood \citep[see Eq.~11 of][]{vincenzi_DES5Ysyst_2024}, 
  3) Supercal-Fragilistic provides the mechanism to create multiple realizations of inter-filter correlated calibrations from the Supercal-Fragilistic covariance matrix for accurate survey simulations, and 
  4) Supercal-Fragilistic is more accurate and precise due to the utilization of more external data. 
The differences in tertiary standard stars between what was used in DES3YR and the Y6 catalog used in this work are $\sim$9~mmag for $g-z$, $\sim$0~mmag for $g-i$, $\sim$5~mmag for $g-r$.

The AB offset uncertainties reported in the C26202-based analysis of \citet{rykoff23} are $\sim$0.011 mag. The reported DES-SN5YR uncertainties (stat+syst) in Supercal-Fragilistic on the diagonal of the covariance matrix are half the size (6~mmag), which is the result of leveraging multiple surveys utilizing multiple primary standard stars. The full Supercal-Fragilistic covariance\footnote{\url{https://github.com/PantheonPlusSH0ES/DataRelease}} is used to determine the effects of correlated systematic uncertainties in both light-curve fitting and in SALT3 model training. Systematic uncertainties due to absolute calibration of the DECam and low-$z$ filters are discussed in Section~6.1 of \citet{vincenzi_DES5Ysyst_2024}.

\subsection{Forced photometry on Image Differences}

The DES difference image pipeline (\DIA{}) obtains a first and preliminary measurement of transient fluxes using point-spread-function (\PSF{}) photometry on images obtained as result of Difference Image Analysis \citep[DIA]{alard_1998_dia,alard_2000_dia}. Details of the DES implementation of difference imaging are provided in \cite{Kessler2015} and are also outlined in Table~\ref{tab:feature_comparison}. In summary, \DIA{} uses a deep template image, constructed from co-adding science verification images obtained under very good observing conditions (low sky-noise and small \PSF{} size). The template image is transformed to match the image properties of the nightly observation, first by astrometric registration and then by convolution with an image kernel. The resulting difference image contains signals only in pixels where there are flux changes, either from real sources or image artifacts, or noise. The pixels with detected sources 3.5$\sigma$ above the sky noise level are evaluated by a separate trained Machine Learning artifact rejection code \citep{Goldstein_2015}.

It is important to note that the detection algorithm was performed on search images processed before the start of the next observing night, and on template images from science verification data. While this pipeline was used to discover all transients included in this data release, and to develop cosmology analysis methods (\SMP{}, K19, \citet{moller_des_snn_2022}, \citet{vincenzi_spectro_2019}, \citet{moller_nohostz_2024}, etc.), it has not been optimized for supernova cosmology. The photometric zeropoints in \DIA{} were calculated from the science verification tertiary standard star catalog and thus do not make use of the updated Y6 catalogs. 

Furthermore, the \DIA{} pipeline has not been optimized to reduce the impact of astrometric and seeing dependent biases. Because photometry of tertiary stars on the search images was performed using Source-Extractor \citep{bertin1996sextractor} automatic-aperture estimator (\texttt{MAG\_AUTO}), subtle magnitude, color, seeing, and airmass dependent biases can arise because PSF fitting is performed on the transient in the difference images. 
Additionally, in the \DIA{} pipeline there is no accounting for stellar proper motions over the course of the 5 year survey which results in mmag photometric biases for tertiary stars with high proper motion.
Finally, the location of each candidate is estimated using an average across all bandpasses, thus ignoring atmospheric effects as a function of airmass and source color \citep*{Brout19smp,leeAcevedo_DCR_2023}. These effects have been improved in the final photometric pipeline that leverages a technique called Scene Modeling Photometry (\SMP{}) and incorporates many of these effects both in the model itself or as corrections. 
\begin{table*}
    \centering
    \caption{Differences between \DIA{} and \SMP{}}
    \begin{tabular}{ccc}
     \toprule
    Stages & \DIA{} & \SMP{} Y5 \\
    \midrule
    Template                     & Science Verification Images         & Any high quality images taken before or after transient \\
    Catalog for Zeropoint        & Science Verification Catalog        & Y6 Forward Model Global Calibration \\
    Photometry for Zeropoint     & Source Extractor \texttt{MAG\_AUTO} & PSF Photometry \\
    Tertiary Star Proper motion  & None                                & Linear fit over 5 Years \\
    Astrometry                   & Science Verification                & Updated in \citet{berstein_2017_desastrometry} \\
    Transient Position           & Forced at average \DIA{} position   & Varied position per filter \\
                                 & across filters                      & (common across all epochs)  \\
    Host Galaxy Profile          & From \DIA{} template                & Forward model fitted per filter \\
    Flux Measurement             & DIA + Forced PSF Photometry         & Forward model Scene + Forced PSF Photometry \\ 
                                 &                                     & but with varied position across all images \\
    \bottomrule
    \vspace{.1in}
    \end{tabular}
    \label{tab:feature_comparison}
\end{table*}

\subsection{Scene Modelling Photometry}

The DES \SMP{} pipeline \citep[B19]{Brout19smp} was first developed for the \DESYt{} cosmology analysis \citep{brout19ana}, and in this work we applied it on the full set of \NLCDIFFIMG{} \DIA{} candidates collected during the 5 years of \DESSN{} operations. The resulting light-curve dataset has been used for the \DESYf{} cosmology analysis \citep{vincenzi_DES5Ysyst_2024, DES_SN5Y_cosmo_2024}. 
Since \SMP{} is a model dependent fit, it is not guaranteed to converge for candidates with multi-season variability and thus the total number of \SMP{}-fitted events that we provide in the data release is \NLCSMP{} (Table~\ref{tab:nevents}). The details of the \SMP{} method are discussed in detail in B19. In summary, \SMP{} simultaneously forward models the DECam images of the transient and its host galaxy while accounting for atmospheric and instrumental effects.

For each transient event `search' image with a candidate detection, the \SMP{} model is convolved with the PSF of each image as determined by PSFex \citep{psfex} in Fourier space and is then resampled to match the pixel grid of the image. This results in a series of `model images' that are compared to the observed DECam images. The time series of DECam images that are used for constraining the transient fluxes are trimmed to span the entire light curve of a transient by using an estimation of the time of peak brightness from the \DIA{} light curve and allowing the transient flux to vary for images that occur within 40 days prior and 300 days after the \DIA{} estimated peak. 
Images taken beyond this range (hereafter referred to as `reference' images) are assumed to have zero transient flux and aid in constraining the degeneracy between a point source at the location of the transient and the underlying galaxy model.

\SMP{} presents several improvements over the \DIA{} pipeline, and these are notably not limited to the methodology. There were also procedural and practical improvements that occurred over the years intervening the two separate efforts to process the data (see Table~\ref{tab:feature_comparison}). 
For \SMP{}, reference and search images were re-processed through the FinalCut program \citep{finalcut} and were scaled to a common zeropoint using the updated DES Y3A2 Standard Bandpasses and tertiary standard stars \citep[see][]{Sevilla_Noarbe_2021}. 
The \SMP{} pipeline also incorporates the updated astrometric solution from \citep{berstein_2017_desastrometry}. 

The reference images for \SMP{} are individual exposures, drawn from the highest quality DES images taken over the course of the entire 5-year survey. While more reference images for \SMP{} is desirable, the \SMP{} method scales with $O(N_{\rm References})$ and for the full 5 year set of images, there are often over 1,000 potential reference images in the Deep Fields. We therefore require $N_{\rm References}=N_{\rm Search}$ and prioritize the templates with the best seeing (FWHM), \PSF{}, and sky level in order to improve the convergence of the \SMP{} galaxy model. 

Unlike \DIA{} (which used \texttt{MAG\_AUTO}), the \SMP{} pipeline does not allow the position of tertiary stars to float on each image, but rather performs forced-position PSF fitting photometry on the tertiary stars after accounting for their proper motions. \SMP{} maintains consistency in the photometric methodology between the tertiary stars and the transient (also forced \PSF{} fitting photometry) and it is this consistency that is essential for mitigating biases in the calibrated transient flux \citep{rest16}. 
Additionally, the \SMP{} pipeline was developed to account for the proper motion of the tertiary stars over the course of the 5-year survey by incorporating a linear fit in RA and DEC to the identified single epoch centroid positions on each night. While the tertiary stars have high signal to noise ratio (SNR) and thus their forced positions can be accurately measured, the transients usually have relatively low SNR and their positions are less certain. Consequently, we incorporate the RA and DEC of each transient candidate into the fitted \SMP{} model. This approach offers the added advantage of the posterior uncertainties naturally accommodating the positional uncertainty of the supernova and accounting for potential photometric bias in the positional uncertainty.

Lastly, it is important to note that the \SMP{} pipeline as implemented by B19 fits an independent model for each bandpass. This mitigates any source color and filter dependent atmospheric effects. Additional atmospheric corrections \citep*[such as in ][]{leeAcevedo_DCR_2023} are minimal (see also Section~\ref{sec:furtherimprovements}).

\begin{figure*}
    \centering
    \includegraphics[width=0.85\linewidth]{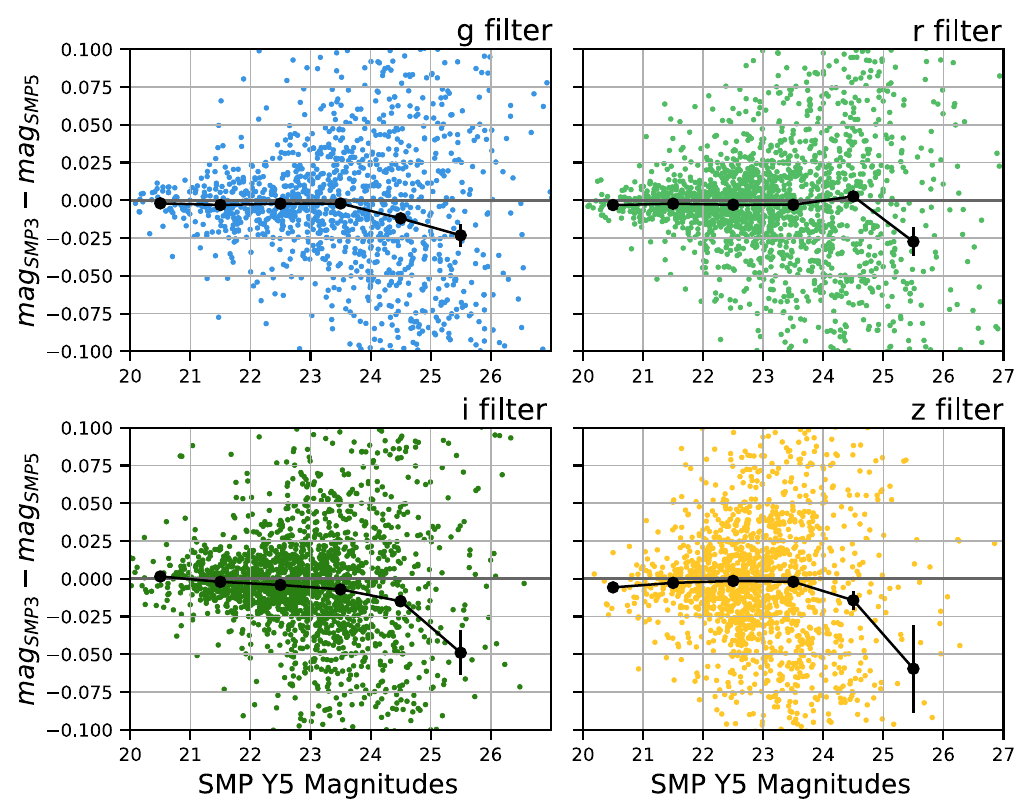}
    \caption{Difference in magnitudes for transients analyzed with \SMP{} Y3 (B19) and \SMP{} Y5 (this work). We expect subtle differences due to the improved catalogs, image processing and astrometry pipelines, and underlying photometric uncertainties. The trend seen in the $i$-band is at $2\sigma$ significance, which is the largest out of all bands.}
    \label{fig:delta_mag_y3y5}
\end{figure*}

We compare the released Year 3 (Y3) photometry in \citet{Brout19smp} with our more recent reprocessing (Y5) of the same SN~Ia sample but with improved templates, astrometry, and tertiary catalogs. Before performing the comparison we apply offset corrections  [-0.006,+0.007,+0.001,-0.005] in corresponding [$g,r,i,z$] filters, to remove changes in the determination of the AB calibration over the time span in which Y3 and Y5 were processed. Fig.~\ref{fig:delta_mag_y3y5} shows the magnitude difference between our Y5 (this work) and the Y3 \citep{Brout19smp} \SMP{} light-curve magnitudes after removing the AB offset differences.
The median is consistent with zero for all filters and the dispersion is found to be the result of the various image processing improvements from Y3-Y5 in combination with independent MCMC fitting. We find for the largest magnitude residuals between Y3-Y5 that these differences are specifically the result of the improved astrometry. The mean uncertainty on $r$-band SN position for the largest residuals ($\Delta{\rm mag}_{\rm SMP}>0.05$) is $0.08$ pixels, which smaller than the average uncertainty of the full sample (0.33 pixels). We observe small offsets in each filter and search for trends as a function of Y5 observed magnitude. Negative residuals for fainter sources are expected with improved low-flux measurements. We find small trends in all bands suggesting improved Y5 sensitivity.

\subsection{Nightly fluxes and their uncertainties}
The DES SN program observed 10 fields in the sky: 8 ``SHALLOW" fields and 2 ``DEEP" fields. Some transients are observed multiple times per night in the same filter. In these cases, we average the results (in DEEP fields and $z$-band, see Tab.~1 from \cite{Kessler2015}). In the case of \DIA{}, the co-addition takes place at the pixel level, by inverse variance weighted image stacking, whereas \SMP{} examines every image as an independent data vector and obtains a flux per individual image, which is later used to calculate the light-curve epoch fluxes using again the inverse variance weighted average.

The photometric uncertainties in the \DIA{} are determined solely from the PSF fit procedure. The \SMP{} photometry flux uncertainties per observation $t_i$ are determined as follows:
\begin{equation}
\sigma_{t_i} = \sqrt{ \sigma_{\rm SMP}^2 + \sigma_{\rm source}^2 + \sigma_{\rm hostgal}^2 }
    \label{eq:sigma_smp}
\end{equation}
where $\sigma_{\rm SMP}$ is determined from the marginalised posterior.
The \SMP{} likelihood (\citealp[B19]{Brout19smp} Eq.~1) contains sky noise but also includes a non-linear component due to uncertainty in the fitted position of the SN. Mean fitted position uncertainties for the SNe are given in Table~\ref{tab:astrometric_variances}. The \SMP{} galaxy model fit from all observations is convolved with the individual night PSF to obtain the Poisson noise contribution of the host galaxy $\sigma_{\rm hostgal}$. The nightly Poisson source noise ($\sigma_{\rm source}$) is included after fits are performed following \citet{astier_photometry_2013}.

\begin{table}
 \caption{Mean uncertainty (in pixels and in arcsec) on the fitted SN position obtained with \SMP{}.}
 \centering
 \begin{tabular}{c|c|c}
 \toprule
Band & $<\sigma_{xy}>$ $\left[\rm{px}\right]$& $<\sigma_{\rm (RA, Dec)}>$ $\left[''\right]$\\
 \midrule
g & 0.43 & 0.12 \\ 
r & 0.33 & 0.09 \\ 
i & 0.31 & 0.08 \\ 
z & 0.32 & 0.09 \\ 
\bottomrule
 \end{tabular}
 \label{tab:astrometric_variances}
\end{table}

For light curve fitting in the cosmology analysis, nightly fluxes are estimated as the variance weighted average $f = \sum_{t_i} f_{t_i} w_{t_i} / \sum_{t_i} w_{t_i}$ with $w_{t_i} = \sigma_{t_i}^{-2}$, thus the total \SMP{} uncertainty per epoch is $\sigma_{\rm stat}^2 = 1/ \sum_{t_i} w_{t_i}$.

\subsection{Host Surface Brightness Anomaly corrections}
The so-called  `surface brightness anomaly' \citep{Kessler2015}, is a systematic underestimation of flux uncertainties for SNe located in high local galaxy surface brightness.

\begin{figure*}
    \centering
    \includegraphics[width=1\linewidth]{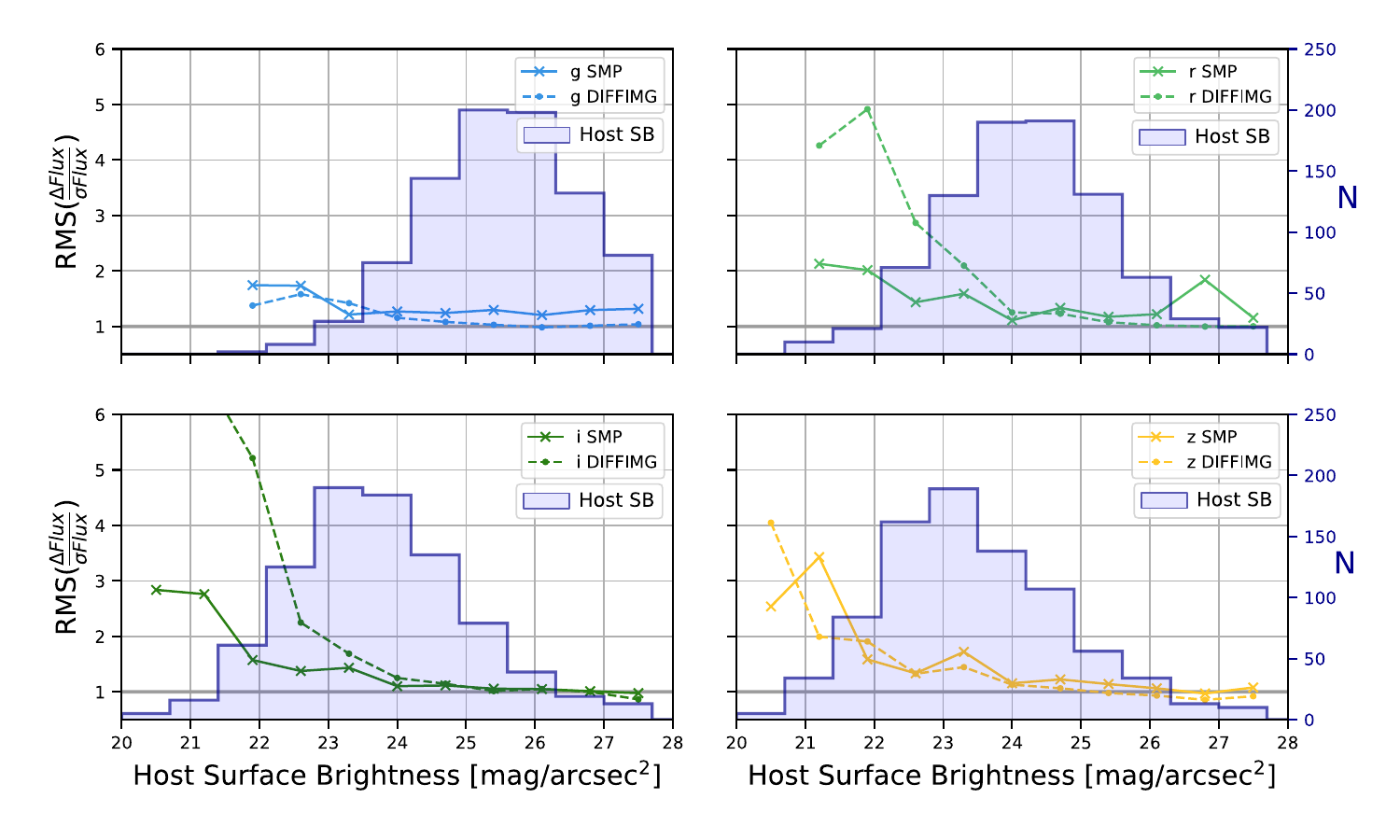}
    \caption{RMS for the flux pull on epochs free of transient flux (assuming true flux of 0), as a function of the Surface Brightness of galaxy host, exactly at the location of the transient. We include the distribution of Host Surface Brightness for each filter in blue. The horizontal line indicates the expected $RMS=1$ for a unit dispersion Gaussian. We show the results for \SMP{} (solid line) and \DIA{} (dashed line).}
    \label{fig:SB_pulls}
\end{figure*}

In \citet{Brout19smp}, it was suggested that this anomaly would be reduced with improved astrometry. We estimate the RMS of flux pulls on epochs free of transient flux (i.e., where true flux is zero) and plot their mean value as a function of the host galaxy surface brightness. In Fig.~\ref{fig:SB_pulls}, we present our results for all bands separately, and for both for \DIA{} 
and \SMP{}. 
We see that for $g$ band the SB anomaly is unimportant in comparison with the redder bandpasses. For $r$ and $i$ bands we find that \SMP{} clearly reduces the excess scatter although (as also in $z$ band) the effect is not fully corrected.
We define a scale correction $S$ as the $\rm{RMS}(\Delta f/\sigma_{\rm stat} )$, and as in B19 we scale the \SMP{} light curve photometric uncertainties $\sigma_f$ as follows
\begin{equation}
\label{eqn:scaleuncertainties}
    \sigma_f = \sigma_{\rm stat} \times S.
\end{equation}

\subsection{Wavelength-dependent atmospheric corrections}
\label{sec:furtherimprovements}
After the \SMP{} pipeline, we perform two sets of corrections to the \DESSN{} photometry.  These corrections are the result of wavelength-dependent atmospheric effects. 

First, the wavelength-dependency of the atmospheric refractive index 
results in Differential Chromatic Refraction \citep[DCR, e.g., ][]{filippenko1982DCR,2009AJ....138...19K}.
Second, atmospheric turbulence causes wavelength-dependent ($\lambda$-dependent) seeing variations. While these photometric biases are similar for both SN and tertiary standard star photometry, they do not cancel out due to the fact that the typical SN spectral energy distribution (SED) is bluer than the typical stellar SED. 

The method to compute the expected corrections to DES-SN photometry is described in \citet*{DES5YR-DCR}. These corrections have not been applied in this data release, rather they are applied as a lookup table in the light-curve fitting process. The lookup table is provided as part of the DES-SN5YR data release (see Appendix~\ref{sec:release}). The DCR and $\lambda$-dependent seeing corrections are small ($\sim3$ mmag) and \citet*{DES5YR-DCR} assess their impact on DES-SN distance estimation and cosmological results to be minimal.

While these corrections could have been included in the \SMP{} model or PSF model, these efforts were developed in parallel for DES and were instead chosen to be included as corrections to the \SMP{} reported fluxes. These corrections are included as supplementary files in the data release as shown in Appendix~\ref{sec:release}. 

\subsection{Comparison of \DIA{} Real-Time Photomery and \SMP{} Photometry}

Training the SALT3 SN~Ia model can be done using public software \citep{SALT3}, and here we compare SALT3 model fits for \DIA{} and \SMP{}, using the training described in \cite{TaylorSALT2vs3}. This training did not include DES data. The number of likely SNe Ia that comprise the `cosmological sample' is \NCosmoSNe{}, which include light curve quality cuts and bias correction viability cuts among others, and it is discussed in great detail in  Table 4 of the companion paper \citet{vincenzi_DES5Ysyst_2024}.

\begin{figure}
    \centering
    \includegraphics[width=1.05\linewidth]{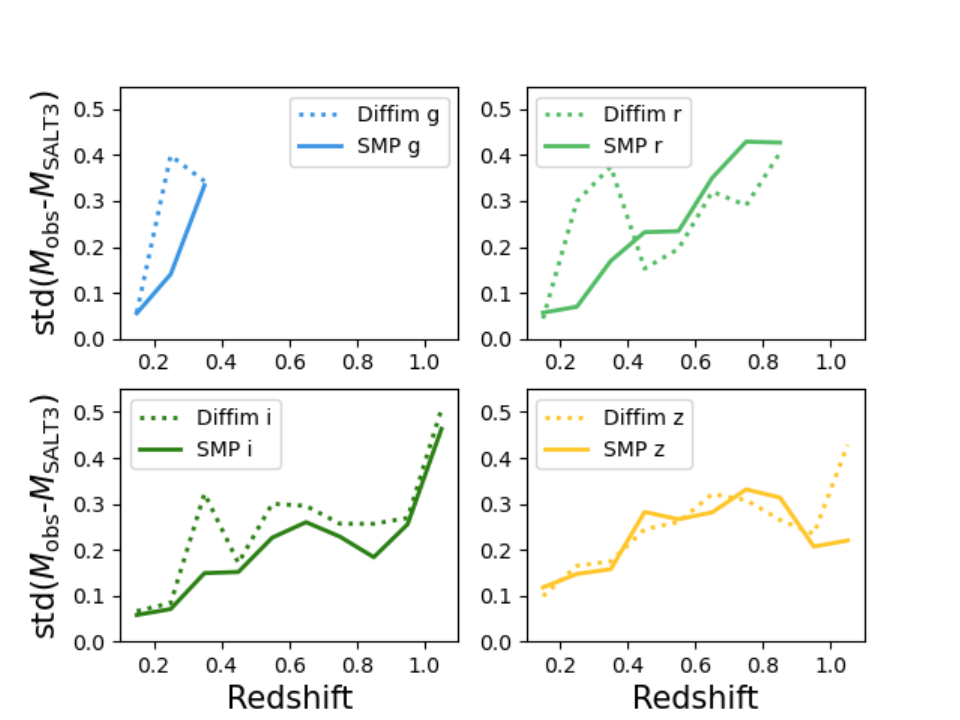}
    \caption{Comparison of flux residuals to SALT3 model fits for \DIA{} and \SMP{} of \DESYf{} light-curves. We generally find either less amount of residual scatter for \SMP{} photometry in transients across the full redshift range.}
    \label{fig:salt3res}
\end{figure}

In Fig.~\ref{fig:salt3res}, we show the light-curve model residuals to this subset of SNe for both \DIA{} and \SMP{}. The scatter of residuals about the SALT3 model is typically reduced in the \SMP{} version of photometry in all bands for a wide range of redshifts (and a wide range of Signal to Noise Ratio SNR values). 

We include the total number of transients with \DIA{} photometry that were run through \SMP{} in Table~\ref{tab:nevents}. We also include the number of light curves with successful \SMP{} chain runs in the 4 DES filters, and those with any 3 succesful filters, 2, and only 1 successful DES filters. We also include the number of events that pass a multi-season transient cut designed to remove AGNs and long-lived transients, and the number of events that have a classification probability of being a SNIa $(P_{\rm SNIa}>0.5)$ obtained with \texttt{SuperNNova} classifier \citep{moller_des_snn_2022}. We note that not all \SMP{} fits converge in all filters. The \SMP{} process relies on several fundamental assumptions that if not satisfied can result in non-convergence of the fits (e.g. the assumption of zero transient flux outside the season of the transient's peak flux). When we select the sample of events that pass the multi-season cut from \citet{moller_des_snn_2022} (i.e.\ are not AGN or long lived transients), we find that for $13,507$ ($96\%$) \SMP{} have successful convergence for at least one filter, and for $13,398$ of these events ($95\%$) \SMP{} processes all 4 filters successfully. The ratio of light curves that have successful \SMP{} convergence in all 4 filters relative to the number of difference imaging candidates is around 50\%, which is consistent with the Monte Carlo simulations done in \cite{Kessler2015} that predicted 75\% of the \DIA{} candidates are not in fact SNe, and 45\% are not single epoch artifacts. When selecting events with probability scores $P_{\rm SNIa} > 0.5$ we find that $4728$ ($92\%$) of the events have successful \SMP{} fits in at least one filter, and $4702$ ($91\%$) events have successful \SMP{} fits in all 4 filters.

In Table~\ref{tab:nevents} we also show the number of events that pass cuts related to the cosmology analysis from \citet{vincenzi_DES5Ysyst_2024}, including events for which DES obtained host galaxy redshift information and also pass the conventional SALT3 model parameter cuts, and final Hubble Diagram (HD) quality cuts. This shows that \SMP{} delivers photometry measurements with higher quality, that ultimately increases the size of the \DESYf{} cosmology sample.

\begin{table*}
    \caption{Number of transients processed with the \DIA{} and \SMP{} pipelines}
\centering
    \begin{tabular}{l|ccccc|cccc}
    \toprule
        $\#$ filters     &           Total &               4 &      3 &     2 &     1 & Multi-season Cut & $P_{\rm SNIa} > 0.5$& W/Host-$z$ \& SALT3 Fit &  HD Quality Cut \\ \midrule
        $\#$ \DIA{} LCs  & $\NLCDIFFIMG{}$ & $\NLCDIFFIMG{}$ & -      & -     & -     &         $14068$&           $5153$&                       $4032$ &                $1499$ \\ \bottomrule
        $\#$ \SMP{} LCs  & $\NLCSMP{}$     & $14486$         & $4179$ & $702$ & $339$ &         $13507$&           $4728$&                       $3621$ &                $1635$ \\ \bottomrule
    \end{tabular}
        \vspace{.1in}
    \label{tab:nevents}
\end{table*}

\section{Auxiliary data}
\subsection{DES-SN Redshifts}
For each SN, we identify the host galaxy using the Directional Light Radius (DLR) method presented by \citet{Sullivan_2010}. The galaxies identified as likely hosts of DES transients are targeted using the AAOmega spectrograph on the 3.9-m Anglo-Australian Telescope (AAT) as part of the OzDES program \citep{2015MNRAS.452.3047Y, 2017MNRAS.472..273C, lidman2020ozdes}. A full description of the different sources of redshifts used in our sample and the spectroscopic efficiency of OzDES+external catalogues are presented by \citet{Vincenzi_2020} and \citet{Brout19smp}.

\subsection{DES-SN Host galaxy properties}
\label{sec:DEShost}
Galaxy properties of DES SN hosts are measured using DES broad-band photometry and, when available, $u$-band data from SDSS and $JHK$ data from VISTA\footnote{The additional $uJHK$ photometry is available for the C3, X3, E2. To evaluate the impact of near infrared data in our fits, we measure host masses and $u$-$r$ rest-frame colours with and without $uJHK$ photometry and do not find any significant bias across the full DES redshift range.} \citep{sutherland_vista_2015}.

DES broad-band photometry is measured from the DEEP coadds presented by \citet{DES_deepstacks}. We use the galaxy SED fitting code by \citet{Sullivan_2010} and the P\'EGASE2  galaxy spectral templates of \citet{1997A&A...326..950F, 2002A&A...386..446L}, assuming a \citet{2001MNRAS.322..231K} initial mass function. The fraction of potentially mismatched SN hosts is modelled and studied in detail by \citet{Qu_hostMismatch} and all identified potential host galaxies are included in the data release.

\subsection{Low-$z$ Redshifts}
For the low-$z$ SNe, we use spectroscopic redshifts revised by \citet{2022PASA...39...46C} and  corrected for peculiar velocities by \citet{2022ApJ...938..112P}. These peculiar velocity corrections are based on 2M++ density fields \citep{2015MNRAS.450..317C} with global parameters found in \citet{2020MNRAS.497.1275S}, combined with group velocities estimated from \citet{2015AJ....149..171T}. We consider uncertainties on peculiar velocity estimates to be 240 km $s^{-1}$.

\subsection{Low-$z$ host galaxy properties}
For the low-redshift sample, we use the same SED fitting code and assumptions implemented to measure host galaxy properties in the DES sample. Low-$z$ optical photometry is combined with near-UV photometry from SDSS $u$-band images and GALEX. This ensures coverage in restframe $u$ band, essential to reliably determine the $u-r$ restframe color. 

We re-measure all the optical and near-UV low-$z$ host galaxy photometry using the open-source package \texttt{hostphot} \citep{2022JOSS....7.4508M}.\footnote{\url{https://github.com/temuller/hostphot}} We measure the global galaxy photometry and visually inspect every SN host galaxy photometric measurement to ensure that the selected aperture is correct.

\section{Conclusion}
We introduce the \DESYf{} light curve sample, obtained with the \SMP{} photometric technique. This sample constitutes the largest collection of SN-like transient light-curves up to $z \sim 1$ obtained with a single instrument. This sample significantly improves on the quality of the photometry measurement with respect to forced-PSF photometry on \DIA{} images by leveraging advances in survey calibration, star proper motion modeling, adequate calibration star PSF photometry, mitigation of atmospheric chromatic effects and source-color effects, among others. We also find a reduction on the surface brightness anomalous scatter effect in \SMP{} with respect to \DIA{} photometry. The light-curve fluxes are consistent with the previous-released \DESYt{} sample, after accounting for zero point offsets.

This data set of Type Ia Supernova light-curves is used in \citet{vincenzi_DES5Ysyst_2024} and \citet{DES_SN5Y_cosmo_2024} to obtain the most accurate constraint on the parameters of the current standard model of cosmology to date. The \DESYf{} provides independent measurement of the accelerated expansion of the universe and paves the way for the next generation (Stage IV) cosmological probes with the Vera Rubin Observatory Legacy Survey of Space and Time and the future space-based Nancy Grace Roman Survey Telescope. We note that the scene modeling pipeline performed here consumed roughly 6 million CPU hours for our 20,000 candidates in DES. Future experiments will need to leverage speed improvements available in recent differentiable models, or GPU computing developments to run on the millions of events expected in the next decade.

\section*{Contribution statements}
Author contributions are as follows. Paper writing: BOS, DB, MV. 
Data processing: BOS, KH, DB, MSa.
Data validation: BOS, DB, RK, MV, MSa. 
Code development: DB, BOS, MSa, RK, PA. 
Contributed to ancillary data products: AM, HQ, MSm, MSu, PW, MSa, RCh, DB, MV, CL, LG, RKe, SH, JL, MA, BOS. 
Working group leadership: TD, DS.
Other DES SNWG people contributed to the interpretation and analysis of the data and provided comments on the manuscript.
The remaining authors have made contributions to this paper that include, but are not limited to, the construction of DECam and other aspects of collecting the data; data processing and calibration; developing broadly used methods, codes, and simulations; running the pipelines and validation tests; and promoting the science analysis.

\section*{Acknowledgements}

Funding for the DES Projects has been provided by the U.S. Department of Energy, the U.S. National Science Foundation, the Ministry of Science and Education of Spain, the Science and Technology Facilities Council of the United Kingdom, the Higher Education Funding Council for England, the National Center for Supercomputing Applications at the University of Illinois at Urbana-Champaign, the Kavli Institute of Cosmological Physics at the University of Chicago, the Center for Cosmology and Astro-Particle Physics at the Ohio State University, the Mitchell Institute for Fundamental Physics and Astronomy at Texas A\&M University, Financiadora de Estudos e Projetos, Funda{\c c}{\~a}o Carlos Chagas Filho de Amparo {\`a} Pesquisa do Estado do Rio de Janeiro, Conselho Nacional de Desenvolvimento Cient{\'i}fico e Tecnol{\'o}gico and the Minist{\'e}rio da Ci{\^e}ncia, Tecnologia e Inova{\c c}{\~a}o, the Deutsche Forschungsgemeinschaft and the Collaborating Institutions in the Dark Energy Survey.

The Collaborating Institutions are Argonne National Laboratory, the University of California at Santa Cruz, the University of Cambridge, Centro de Investigaciones Energ{\'e}ticas, Medioambientales y Tecnol{\'o}gicas-Madrid, the University of Chicago, University College London, the DES-Brazil Consortium, the University of Edinburgh, the Eidgen{\"o}ssische Technische Hochschule (ETH) Z{\"u}rich, Fermi National Accelerator Laboratory, the University of Illinois at Urbana-Champaign, the Institut de Ci{\`e}ncies de l'Espai (IEEC/CSIC), the Institut de F{\'i}sica d'Altes Energies, Lawrence Berkeley National Laboratory, the Ludwig-Maximilians Universit{\"a}t M{\"u}nchen and the associated Excellence Cluster Universe, the University of Michigan, NSF's NOIRLab, the University of Nottingham, The Ohio State University, the University of Pennsylvania, the University of Portsmouth, SLAC National Accelerator Laboratory, Stanford University, the University of Sussex, Texas A\&M University, and the OzDES Membership Consortium.

Based in part on observations at Cerro Tololo Inter-American Observatory at NSF's NOIRLab (NOIRLab Prop. ID 2012B-0001; PI: J. Frieman), which is managed by the Association of Universities for Research in Astronomy (AURA) under a cooperative agreement with the National Science Foundation.

Based in part on data acquired at the Anglo-Australian Telescope. We acknowledge the traditional custodians of the land on which the AAT stands, the Gamilaraay people, and pay our respects to elders past and present. Parts of this research were supported by the Australian Research Council, through project numbers CE110001020, FL180100168 and DE230100055. Based in part on observations obtained at the international Gemini Observatory, a program of NSF’s NOIRLab, which is managed by the Association of Universities for Research in Astronomy (AURA) under a cooperative agreement with the National Science Foundation on behalf of the Gemini Observatory partnership: the National Science Foundation (United States), National Research Council (Canada), Agencia Nacional de Investigaci\'{o}n y Desarrollo (Chile), Ministerio de Ciencia, Tecnolog\'{i}a e Innovaci\'{o}n (Argentina), Minist\'{e}rio da Ci\^{e}ncia, Tecnologia, Inova\c{c}\~{o}es e Comunica\c{c}\~{o}es (Brazil), and Korea Astronomy and Space Science Institute (Republic of Korea).  This includes data from programs (GN-2015B-Q-10, GN-2016B-LP-10, GN-2017B-LP-10, GS-2013B-Q-45, GS-2015B-Q-7, GS-2016B-LP-10, GS-2016B-Q-41, and GS-2017B-LP-10; PI Foley).  Some of the data presented herein were obtained at Keck Observatory, which is a private 501(c)3 non-profit organization operated as a scientific partnership among the California Institute of Technology, the University of California, and the National Aeronautics and Space Administration (PIs Foley, Kirshner, and Nugent). The Observatory was made possible by the generous financial support of the W.~M.~Keck Foundation.  This paper includes results based on data gathered with the 6.5 meter Magellan Telescopes located at Las Campanas Observatory, Chile (PI Foley), and the Southern African Large Telescope (SALT) (PIs M.~Smith \& E.~Kasai).

The DES data management system is supported by the National Science Foundation under Grant Numbers AST-1138766 and AST-1536171. The DES participants from Spanish institutions are partially supported by MICINN under grants ESP2017-89838, PGC2018-094773, PGC2018-102021, SEV-2016-0588, SEV-2016-0597, and MDM-2015-0509, some of which include ERDF funds from the European Union. IFAE is partially funded by the CERCA program of the Generalitat de Catalunya. Research leading to these results has received funding from the European Research Council under the European Union's Seventh Framework Program (FP7/2007-2013) including ERC grant agreements 240672, 291329, and 306478. We  acknowledge support from the Brazilian Instituto Nacional de Ci\^encia e Tecnologia (INCT) do e-Universo (CNPq grant 465376/2014-2).

This manuscript has been authored by Fermi Research Alliance, LLC under Contract No. DE-AC02-07CH11359 with the U.S. Department of Energy, Office of Science, Office of High Energy Physics.

This research used resources of the National Energy Research Scientific Computing Center (NERSC), a U.S. Department of Energy Office of Science User Facility located at Lawrence Berkeley National Laboratory, operated under Contract No. DE-AC02-05CH11231 using NERSC award HEP-ERCAP0023923.

We acknowledge the University of Chicago’s Research Computing Center for their support of this work.

\newpage
\appendix

\section{Data Release Details}
\label{sec:release}
We provide difference imaging photometry for \NLCDIFFIMG{} candidates, \SMP{} photometry for \NLCSMP{} candidates, and ancillary data at \DESDRLINK{} 

\vspace{.1in}
For the \SMP{} light curves, we provide the metadata: 

\begin{itemize}
\setlength\itemsep{.01em}

\item \texttt{REDSHIFT} - determined from the host galaxy using OzDES and other external catalogues
\item \texttt{HOST\_g/r/i/z} - determined from deep co-added DES images \citep{DES_deepstacks} 
\item \texttt{HOST\_LOGMASS} and \texttt{HOST\_COLOR} - Host stellar mass and $u$-$r$ rest-frame color, determined as described in Sec.~\ref{sec:DEShost}
\end{itemize}

\vspace{.1in}
We also provide the following time series data:
\begin{itemize}

\setlength\itemsep{.01em}

\item \texttt{MJD} - Modified Julian Date of Light Curve Data
\item \texttt{ZPT} - Re-determination of Zeropoint of Image for \SMP{} scaling
\item \texttt{BAND} - DES filter 
\item \texttt{SKY} - sky level in units of image counts
\item \texttt{PSF} - FWHM of PSF in units of arcsec
\item \texttt{FLUXCAL} - Reported \SMP{} flux in units of \textit{fluxcal counts} (fixed standard zeropoint 27.5)
\item \texttt{FLUXCALERR} - Reported \SMP{} flux uncertainty in units of \textit{fluxcal counts}
\end{itemize}

\vspace{.1in}

For the low-$z$ sample, we release the updated host galaxy photometry and properties:
\begin{itemize}

\setlength\itemsep{.01em}

\item \texttt{HOST\_NUV/u/g/r/i/z} - determined from \citet{DES_deepstacks}
\item \texttt{HOST\_LOGMASS} and \texttt{HOST\_COLOR} - Host stellar mass and $u$-$r$ rest-frame color, determined as described in Sec.~\ref{sec:DEShost}
\end{itemize}

\vspace{.1in}
Additional corrections:
\begin{itemize}
\item DCR Correction \citep*{leeAcevedo_DCR_2023} lookup table
\end{itemize}

\newpage
\subsection{Light curve example}

An example light curve in text format is included for general reference.

\begin{figure*}[!ht]
\centering
\tiny
\begin{minipage}{0.9\textwidth}
\begin{lstlisting}
SURVEY:               DES
SNID:                 1246273
SNTYPE:               0
FAKE:                 0
FILTERS:              griz
NXPIX:                2048
NYPIX:                4096
PIXSIZE:              0.263000
RA:                   54.566567
DEC:                  -27.994892
REDSHIFT_HELIO:        2.352000 +-  0.001000
FIELD:                C1

HOSTGAL_NMATCH:       1
HOSTGAL_NMATCH2:      1
HOSTGAL_OBJID:        590
HOSTGAL_SPECZ:         2.352000 +-  0.001000
HOSTGAL_SNSEP:        0.0729999989271164
HOSTGAL_DDLR:         0.11299999803304672
HOSTGAL_RA:           54.566578
HOSTGAL_DEC:          -27.994910
HOSTGAL_MAG: 22.800 22.830 22.600 22.110
HOSTGAL_MAGERR: 0.010 0.010 0.010 0.010
HOSTGAL_SB_FLUXCAL: 15.530 12.930 18.200 29.040

# computed quantities
REDSHIFT_CMB:          2.350790 +-  0.001000
MWEBV:                 0.010100 +-  0.001700
PEAKMJD:              56552.26953125
MJD_DETECT_FIRST:     56538.365000
MJD_DETECT_LAST:      -9.0

HOSTGAL_PHOTOZ:       -999 +- -999
HOSTGAL_LOGMASS:      10.656000 +-  0.019000
HOSTGAL2_PHOTOZ:      -999 +- -999
HOSTGAL2_LOGMASS:     -999 +- -9
VPEC:                  0.000000 +- 300.000000

# PRIVATE (non-standard) variables
PRIVATE(DES_numepochs_ml_Y1):  28
PRIVATE(DES_numepochs_ml_Y2):  0
PRIVATE(DES_numepochs_ml_Y3):  12
PRIVATE(DES_numepochs_ml_Y4):  32
PRIVATE(DES_numepochs_ml_Y5):  16
PRIVATE(AGN_SCAN):    -1

# sim/truth quantities
SIM_TYPE_INDEX:       -9

# --------------------------------------
# obs info
NOBS: 71
NVAR: 14

VARLIST: MJD BAND FIELD PHOTFLAG XPIX YPIX CCDNUM IMGNUM GAIN FLUXCAL FLUXCALERR PSF_SIG1 ZEROPT SKY_SIG
OBS: 56534.2230 i  C1  14336  776.9 1361.6  62 228748  3.770   5.4924e+01   8.7440e+00   2.2620 32.280  20.30
OBS: 56534.2250 z  C1  14336  779.5 1355.6  62 228749  4.050   8.2914e+01   1.0723e+01   2.2470 32.899  42.80
OBS: 56538.3700 i  C1  13312  778.5 1291.1  62 230181  3.740   7.1572e+01   6.8920e+00   2.5070 32.601  20.80
OBS: 56538.3720 z  C1  13312  786.0 1295.1  62 230182  4.080   7.5552e+01   8.2890e+00   2.4130 33.032  38.40
...
OBS: 56607.0620 z  C1  12288  710.3 1289.3  62 252746  4.050   5.8884e+01   1.4748e+01   3.4040 32.928  45.60
OBS: 56614.0730 g  C1  12288  724.3 1326.2  62 255449  4.380   6.9555e+01   1.9029e+01   1.9910 32.223  38.40
OBS: 56614.0780 i  C1  12288  710.0 1313.5  62 255451  3.900   5.1747e+01   8.9140e+00   1.6640 32.579  39.60
OBS: 56614.0800 z  C1  12288  718.3 1310.7  62 255452  4.070   6.8474e+01   8.1760e+00   1.5850 33.046  58.20
OBS: 56615.0960 g  C1  12288  726.1 1342.4  62 255888  4.330   5.3188e+01   1.9439e+01   2.1090 32.015  36.80
OBS: 56615.1010 i  C1  12288  711.0 1360.6  62 255890  3.830   5.2356e+01   9.5250e+00   1.8340 32.560  36.20
OBS: 56615.1040 z  C1  12288  718.8 1367.2  62 255891  4.020   5.6826e+01   1.1180e+01   1.9620 32.996  52.80
OBS: 56628.0800 i  C1  12288  718.9 1312.5  62 259320  3.820   3.6911e+01   6.0560e+00   2.1850 32.554  21.00
OBS: 56628.0830 z  C1  12288  719.6 1304.5  62 259321  4.060   4.8195e+01   8.7390e+00   2.3730 33.042  41.30
OBS: 56635.1270 g  C1  12288  735.5 1342.8  62 261961  4.230   5.3322e+01   5.0210e+00   2.6680 32.353  11.10
OBS: 56635.1320 i  C1  12288  734.6 1360.7  62 261963  3.780   3.1282e+01   6.5190e+00   2.1480 32.605  24.40
OBS: 56635.1350 z  C1  12288  726.1 1331.3  62 261964  4.060   5.2199e+01   7.6440e+00   1.8760 33.084  49.10
OBS: 56642.0880 i  C1      1  785.9 1334.0  62 265092  3.360   1.5130e+02   5.8947e+01   2.3770 32.201  70.80
OBS: 56642.0900 z  C1      1  770.6 1324.0  62 265093  4.020  -1.9340e+00   1.8025e+01   2.0710 32.776  72.00
OBS: 56645.0780 g  C1  12288  718.7 1326.5  62 266113  4.320   4.9523e+01   1.0656e+01   2.3580 32.331  24.30
OBS: 56645.0820 i  C1  12288  736.7 1331.3  62 266115  3.840   4.8745e+01   7.5700e+00   1.9830 32.574  29.70
...
OBS: 56693.0340 g  C1  12288  736.9 1328.8  62 281587  4.260   2.1540e+01   5.0830e+00   2.1580 32.253  13.00
OBS: 56693.0390 i  C1  12288  749.5 1346.6  62 281589  3.780   2.6410e+01   6.3810e+00   1.8870 32.529  24.90
OBS: 56693.0410 z  C1  12288  742.8 1354.2  62 281590  4.070   3.3742e+01   7.8600e+00   1.8980 32.982  44.70
END:
\end{lstlisting}
\end{minipage}
\end{figure*}

\newpage
\subsection{Structure of data release}

The structure of the full data release contents is summarized in Table~\ref{tab:drcontents}.

\begin{deluxetable}{p{0.16\textwidth}p{0.72\textwidth}}[!h]
\centering
\tablecolumns{2}
\tablewidth{20pc}
\tablecaption{Structure of the \DESYf{} Data Release \label{tab:drcontents}} 

\tablehead {\colhead {\textbf{Folder}}  & \colhead {\textbf{Description}}} 
\startdata
    \texttt{0\_DATA} & \DESYf{} light-curves from this work, for all SN candidates that have a spectroscopic redshift and pass light-curve quality cuts.\\
\hline
    \texttt{1\_SIMULATIONS}  & Set of 25 SNANA simulations with the same properties as the DES SN sample, used in the analysis for testing and cosmological analysis validation. \\ 
\hline
    \texttt{2\_LCFIT\_MODEL}   & SALT3 light-curve model SED time-series, used for fitting the DES-SN and Low-$z$ data and all simulated samples.\\ 
\hline
    \texttt{3\_CLASSIFICATION}  & Classification probabilities from the various classification algorithms used in the analysis. These include SuperNNova, SCONE and SNIRF classification for the Hubble Diagram events and SuperNNova classifications from \cite{moller_nohostz_2024}.\\  
\hline
    \texttt{4\_DISTANCES\_COVMAT} & SN distance moduli measured after bias-corrections and correcting for contamination. This includes all the Beams with Bias Correction (BBC) input files used to reproduce the Hubble diagram. $N_{\rm SN}$-dimensional systematic covariance matrices (statistical only and statistical+systematic for all systematics combined, and also individual systematic matrices).\\ 
\hline
    \texttt{6\_DCR\_CORRECTIONS} & Wavelength-dependent Photometric Corrections. \\
\hline
    \texttt{5\_COSMOLOGY}  & Chains and resulting cosmological constraints for the different models presented in \cite{DES_SN5Y_cosmo_2024}. \\
\hline
    \texttt{7\_PIPPIN\_FILES} & This folder includes the Pippin input files needed to reproduce DES simulations and cosmological analysis.
\enddata
\end{deluxetable}

\bibliography{biblio}
\bibliographystyle{yahapj_twoauthor_amp} 

\end{document}